# CVD of $CrO_2$: towards a lower temperature deposition process

**P.M. Sousa[1], S.A. Dias[1], A.J. Silvestre[2], O. Conde[1*], B. Morris[3], K.A. Yates[3],**

**W.R. Branford[3,4] and L.F. Cohen[3]**

[1]*Faculty of Sciences of the University of Lisbon, Department of Physics and ICEMS*
*Campo Grande, Ed. C8, 1749-016 Lisboa, PT*

[2]*Instituto Superior de Engenharia de Lisboa and ICEMS*
*R. Conselheiro Emídio Navarro 1, 1959-007 Lisboa, PT*

[3]*Imperial College, Blackett Laboratory*
*Prince Consort Road, London SW7 2AZ, UK*

[4]*University College London, Department of Chemistry*
*Gordon Street, London WC1H OAJ, UK*

The ability to manipulate and amplify currents of different spin-type is the basis of an important and rapidly developing field of fundamental and applied research known as spintronics. Among the magnetic materials being actively investigated for their potential use in spintronic devices, chromium dioxide ($CrO_2$) is one of the most attractive because it is a half-metal fully spin polarized at the Fermi level with a Curie temperature above room temperature ($T_c$=393 K) and a calculated magnetic moment of 2 $\mu_B$/f.u..[1-4] Therefore, much effort has been put into developing efficient and controlled methods for preparing $CrO_2$ films. Nevertheless, the synthesis of $CrO_2$ films has been a difficult task because it lacks a thermodynamic stable plateau at atmospheric pressure and, if heated, easily decomposes into the insulating antiferromagnetic $Cr_2O_3$ phase which is the most stable chromium oxide at ambient conditions.

Although efforts to grow $CrO_2$ thin films have involved several techniques,[5-11] until now chemical vapour deposition (CVD) seems to be the most successful one to produce these films. Highly oriented thin films of $CrO_2$ are currently grown on $TiO_2$ rutile phase and $Al_2O_3$(0001) by thermal CVD using $CrO_3$ as chromium precursor.[12-15] $CrO_3$ is vaporised at 260 ºC, carried into the reaction zone by an oxygen/argon flux and thermally decomposed onto a heated substrate. Besides $CrO_3$, other precursors such as $Cr_8O_{21}$[16] and $CrO_2Cl_2$[17,18] have also been used for the CVD of $CrO_2$ layers with similar properties to those grown with $CrO_3$. The quality of the $CrO_2$ films has been claimed to depend critically on the substrate temperature, no matter the precursor used, the best results having been obtained at substrate temperatures between 390 and 400 ºC.[12,14] Up to now, films produced outside this temperature range have always been reported as containing also the $Cr_2O_3$ or $Cr_2O_5$ phases.

*Corresponding author
Olinda Conde
Tel.: +351 217500035, Fax.: +351 217500977
E-mail: oconde@fc.ul.pt

In this work we report on the synthesis of highly oriented a-axis $CrO_2$ films onto $Al_2O_3$(0001) by atmospheric pressure CVD at temperatures as low as 330 ºC. Our experiments were based on the method described by Ishibashi *et al.*[12] and already modified by S.M. Watts.[16,19] Therefore, a quartz tube placed inside a single-zone furnace and an independent control of the substrate temperature were used. However, the reactor was designed with a configuration that attempts to minimize the turbulence along the tube, guaranteeing processing conditions close to the ideal for laminar flux over the substrate surface and increasing the reactor efficiency (see experimental section). $CrO_3$ powder was used as precursor and $O_2$ as carrier gas. The powder was loaded into a stainless steel boat aligned with the substrate holder, 15 cm apart from it. Films were grown for deposition times ($t_{dep}$) of 4 and 8 hours, by using an oxygen flow rate ($\phi_{O2}$) of 50 sccm, a precursor temperature ($T_p$) of 260 ºC and substrate temperatures ($T_s$) ranging from 330 to 410 ºC.

All the films were produced on 10×10 mm$^2$ substrates, are homogeneous and exhibit a good adherence and a black shiny metallic colour characteristic of chromium dioxide. Surface morphology and roughness (rms) depend mainly on film thickness, the lowest value for the rms having been measured ~7 nm over an area of 20×20 μm$^2$ by AFM of 200 nm thick films. Film thickness strongly depends on substrate temperature, values of 1424±49 nm and 57±13 nm having been measured for films deposited during 4 hours at $T_s$=410 ºC and $T_s$=330 ºC, respectively. Therefore, $CrO_2$ deposition rates ($r_{CrO_2}$) varying between 59.3 and 2.4 Å min$^{-1}$ were calculated, the highest values being much larger than those reported in literature [16,17]. The Arrhenius diagram shown in figure 1 allows to determine an apparent activation energy of 121.0±4.3 kJ mol$^{-1}$, indicating that surface chemical reaction kinetics is the rate-limiting step for film growth.

Figure 2 displays the X-ray diffraction (XRD) patterns of films grown at different substrate temperatures, between 330 and 390 ºC. Only two regions of interest are shown, i.e. 39º– 43º and 86º–94º, corresponding to the 2θ ranges where diffraction peaks are observed in the full 25º–95º analysed range. As can be seen, only the (200) and (400) diffraction lines of $CrO_2$ can be clearly identified, besides those from the sapphire substrate indicated by "S", revealing a highly a-axis oriented film growth. Furthermore, the diffraction lines originating from the (006) and (0012) planes of $Cr_2O_3$, at 2θ = 39.7º and 85.7º, which were clearly visible in a previous work[20] carried out in a less efficient CVD reactor, are now absent or at least their intensity is negligible, even for the sample grown at the lowest temperature of 330ºC. In order to assess the degree of orientation of the crystallites, rocking curves (RC) for the (200) reflection were measured and their full-width at half-maximum (FWHM) values plotted as a function of deposition temperature (Fig. 3). As expected, crystal quality improves as growth temperature increases revealed by the narrower of the RC at higher $T_s$; however, for the lowest $T_s$=330ºC the RC-FWHM value is ~0.68º showing that this sample is still strongly textured although it exists some degree of misorientation between the crystallites.

Low temperature grown samples keep the same room temperature magnetic behaviour as those films grown at higher temperatures. Figure 4 shows the room temperature out-of-plane magnetization *vs*. field loops for samples deposited at 380ºC, (674±40 nm), 340ºC (113±22 nm) and 330ºC (62±17 nm), which have respective room temperature saturation magnetizations of 1.30±0.07 $\mu_B$/f.u., 1.5±0.3 $\mu_B$/f.u., 1.5±0.4 $\mu_B$/f.u. The uncertainty is in the magnetic volume. The measured magnetic properties of the films are consistent, within error, with both bulk[5] and high quality film[21] saturation magnetization of 2.0 $\mu_B$/f.u. at 5K and a Curie temperature of 393K, resulting in a room temperature saturation magnetization of ~1.5 $\mu_B$/f.u. for both bulk[5] and films[21]. The inset to figure 4 shows a simple fit to a point contact Andreev reflection spectrum of a junction between a lead tip and the 340ºC film, using the Blonder-Tinkham-Klapwijk (BTK)[22] model modified by Mazin et al.[23] in the diffusive regime, showing that the best fit gives a transport spin polarization of 85%±3%.

In conclusion, we succeeded in lowering the growth temperature of $CrO_2$ films by 70 ºC in relation to published data, which should be looked as a promising result in view of the use of these films with thermally sensitive materials such as those envisaged in spintronic devices (e.g. narrow band gap semiconductors). It was shown that highly oriented a-axis $CrO_2$ films can be synthesised onto $Al_2O_3$(0001) by atmospheric pressure CVD at temperatures as low as 330 ºC. Deposition rates strongly depend on the substrate temperature whereas for film surface microstructures the dependence is mainly on film thickness. For the experimental conditions used in this work, $CrO_2$ growth kinetics is dominated by a surface reaction mechanism with an apparent activation energy of 121.0±4.3 kJ $mol^{-1}$. The magnitude and temperature dependence of the saturation magnetization up to room temperature is consistent with bulk measurements.

## *Experimental*

***Reactor characteristics***: As previously described, our system is similar to that described in ref. 16 and 19, although implementing two significant modifications: *i)* the quartz tube length was shortened to 55 cm reducing the reactor cold zones outside the furnace and, consequently, minimising convective phenomena along the reactor, and *ii)* the distance between the precursor boat and the substrate heather block was increased in order to homogenise the gas mixture, at the substrate surface, that results from the intersection of the oxygen longitudinal flux with the upstream flux of the $CrO_3$ precursor at the boat level. In principle, the process may be scaled up if both phenomena described above are taken into account.

***CVD procedures:*** $CrO_3$ powder (purity 99.9%) was used as precursor and loaded into a stainless steel boat aligned with the substrate holder, 15 cm apart from it. Prior to their insertion into the reactor, the substrates were ultrasonically cleaned in organic solvents, rinsed in distilled water and dried with a $N_2$ flux. Oxygen (purity 99.999%) was used as carrier gas. Films were grown for deposition times of 4 and 8 hours, by using an oxygen flow rate of 50 sccm, a precursor temperature of 260 ºC and substrate temperatures ranging from 330 ºC to 410 ºC. In order to avoid the deposition of any kind of impurities during the initial stages of the deposition process, the substrate was always heated up to the deposition temperature before the melting of the $CrO_3$ precursor occurs (196 ºC).

***Film analysis:*** The morphology and microstructure of the films were analysed by scanning electron microscopy (SEM) and atomic force microscopy (AFM). Film thicknesses were evaluated by image digital processing of SEM cross-section micrographs. Their crystallographic structure was studied by Bragg-Brentano X-ray diffraction (XRD) with Cu K$\alpha$ radiation. Magnetization measurements were performed on an Oxford Instruments vibrating sample magnetometer, with the applied field perpendicular to the surface of the film. Point contact Andreev reflection spectra (which is a measure of the differential conductance $G = dI/dV$ as a function of bias voltage V) were measured on the film grown at 340$^{o}$C using a mechanically sharpened lead tip at 4.2 K. We present the simple analysis of this data using the modified BTK model of Mazin[23] in the diffusive regime. The parameters extracted from that model are defined in ref 22. A more thorough analysis comparing ballistic and diffusive limit fitting to a complete set of Andreev reflection spectra as a function of temperature using both lead and niobium tips will be presented in a future publication. Nevertheless the values for the spin polarisation of the transport current are similar to values reported in the literature for films grown at higher temperatures. We can say that the lower growth temperature has not compromised this important parameter in any obvious respect.

**Acknowledgements**

The authors gratefully acknowledge the financial support of Fundação para a Ciência e Tecnologia (FCT) under contract POCTI/CTM/41413/2001 and the EPSRC grant number GR/T03802. P.M. Sousa also acknowledges FCT for a PhD grant.

## FIGURE CAPTIONS

**Figure 1** - Logarithm of the apparent deposition rate as a function of the substrate reciprocal temperature. ■, measured values; —, Arrhenius line fitting.

**Figure 2** - XRD diffractograms of $CrO_2$ films deposited at different substrate temperatures. All the films were grown for $t_{dep}$= 4 hours except for the film deposited at $T_s$=330 ºC where $t_{dep}$= 8 hours. Film thicknesses are 57, 113, 169, 272, 326, 657 and 846 nm, from 330 ºC up to 390 ºC.

**Figure 3** - Full-Width Half-Maximum (FWHM) of the rocking curves measured for the (200) diffraction peak vs substrate temperature. The inset shows the rocking curve of the film grown at $T_s$=330 ºC for 8 hours.

**Figure 4** - Out-of-plane magnetization *vs*. applied field at room temperature, for samples grown at 380, 340, and 330 ºC. Inset: Point Contact Andreev reflection spectrum (open circles) for 340°C film with junction resistance of 79Ω where G is the differential conductance (dI/dV). The fit to the modified BTK model[22] in the diffusive regime is shown. The parameters from the fit are the polarisation P, the interface parameter Z, the superconducting energy gap Δ and the spreading parameter ω.

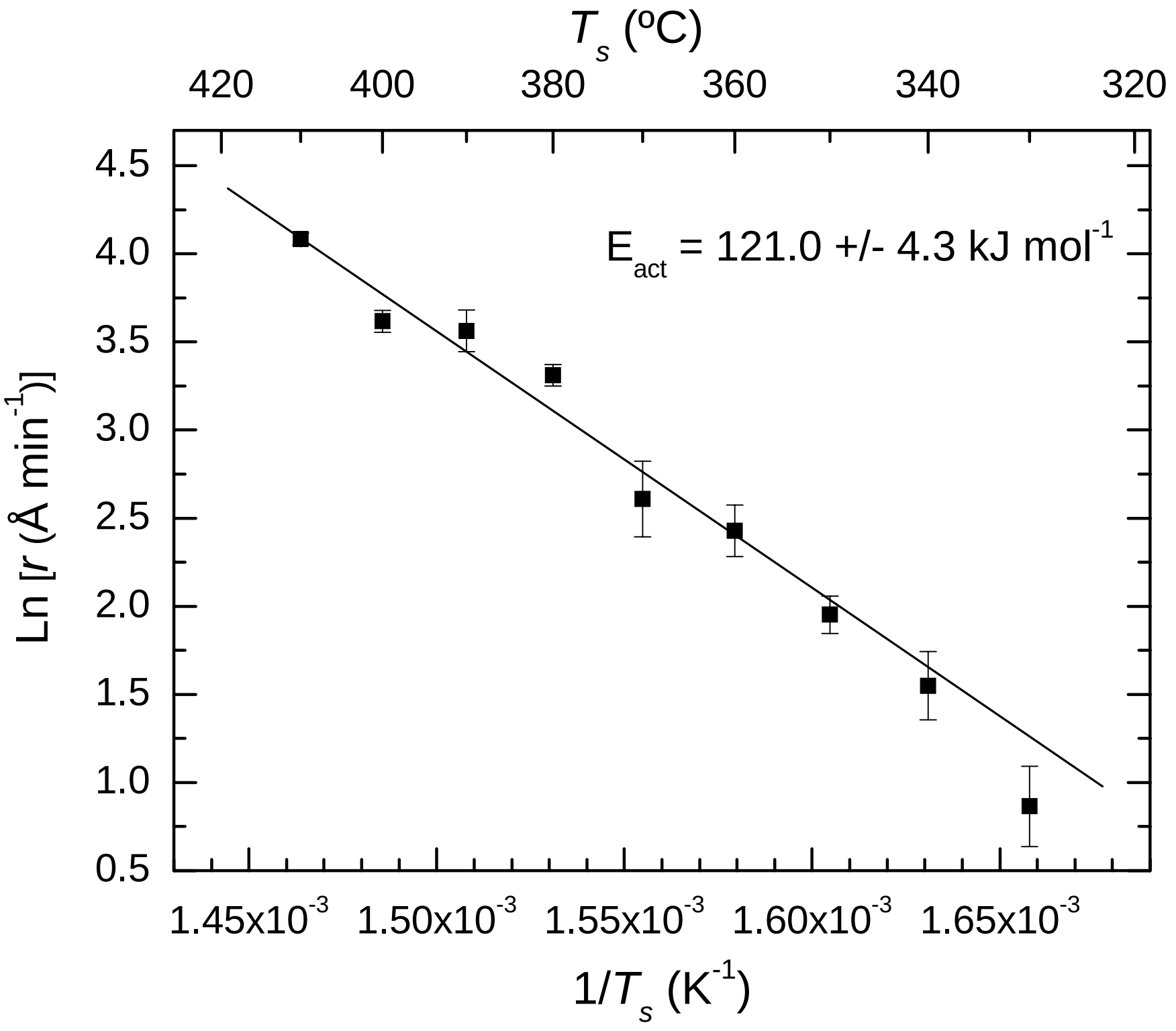

**Figure 1**

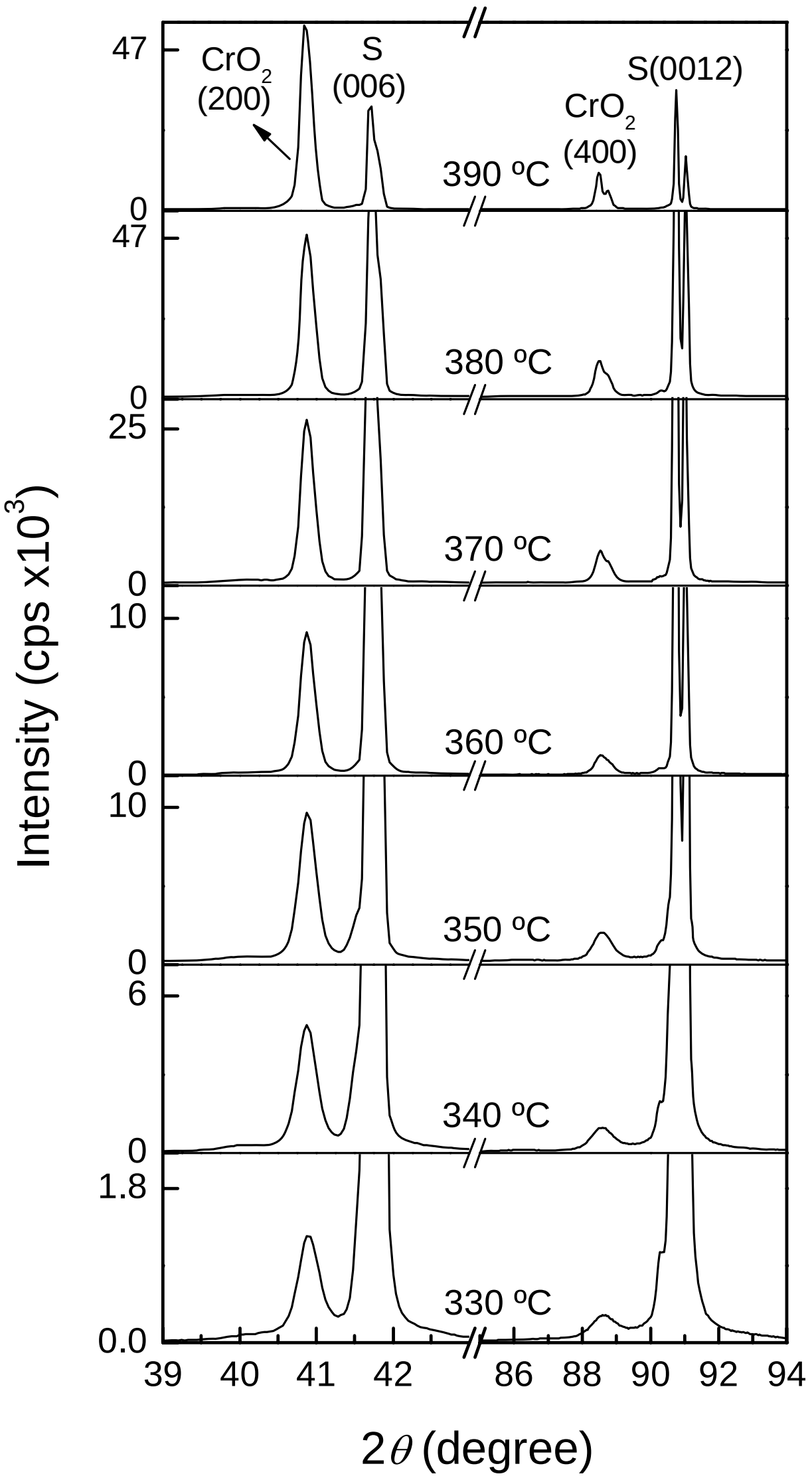
CrO2
(200)
S
(006)
S(0012)
CrO2
(400)
390 ºC
380 ºC
370 ºC
360 ºC
350 ºC
340 ºC
330 ºC
47
0
47
0
25
0
10
0
10
0
6
0
1.8
0.0
39 40 41 42 86 88 90 92 94
Intensity (cps x10^3)
2θ (degree)

**Figure 2**

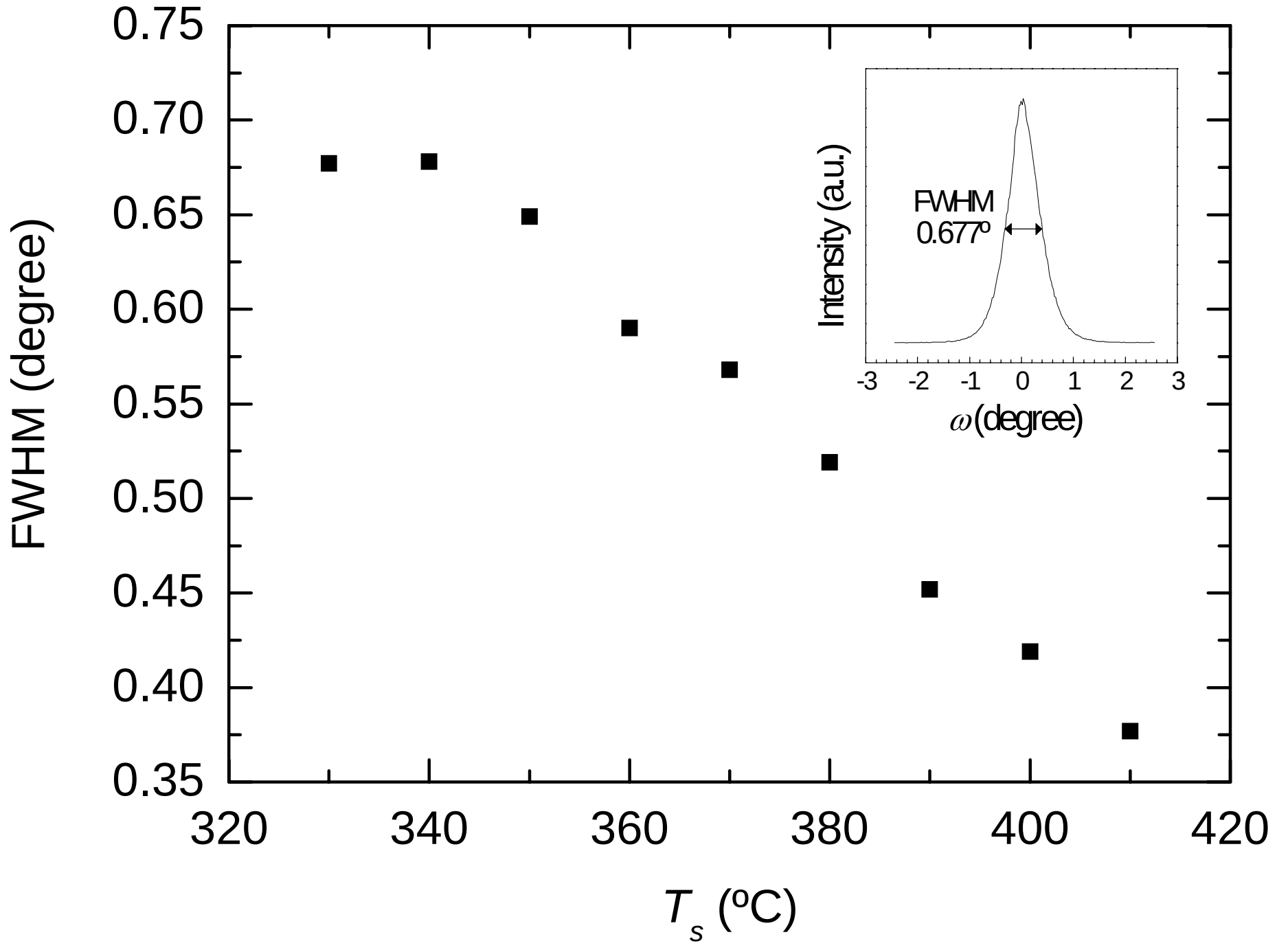

**Figure 3**

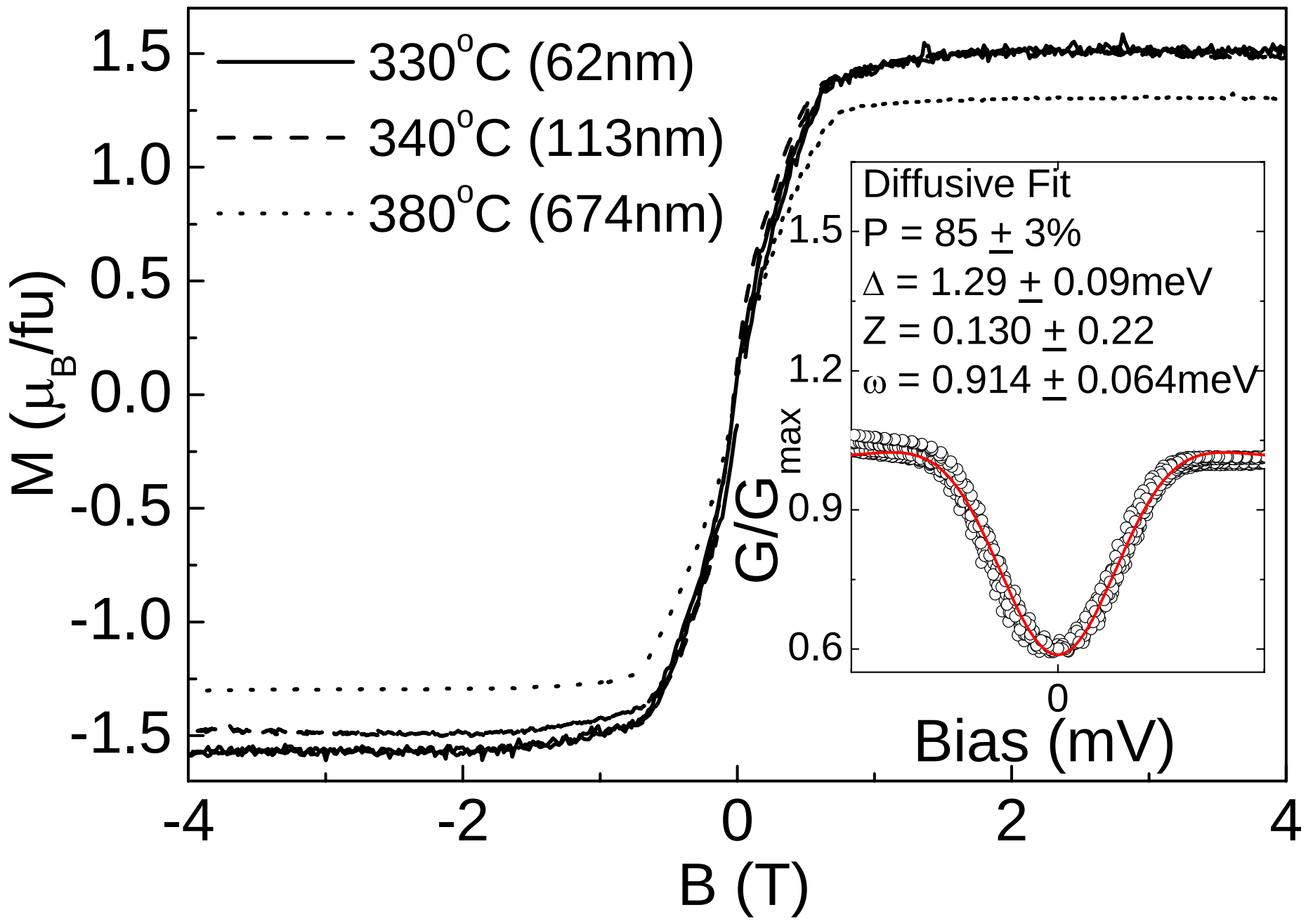
330°C (62nm)
340°C (113nm)
380°C (674nm)
M (μB/fu)
B (T)
Diffusive Fit
P = 85 ± 3%
Δ = 1.29 ± 0.09meV
Z = 0.130 ± 0.22
ω = 0.914 ± 0.064meV
G/Gmax
Bias (mV)

**Figure 4**